# Analyzing Growth Kinematics and Fractal Dimensions of Molybdenum Disulfide Films

Yan Jiang,*,† Moritz to Baben,‡ Yuankun Lin,† Chris Littler,† A.J. Syllaios,†

Arup Neogi,† and Usha Philipose†

†*Department of Physics, University of North Texas, Denton, TX 76203, USA*

‡*GTT-Technologies, 52134 Herzogenrath, Germany*

E-mail: yanjiang@my.unt.edu

**Abstract**

Through the positive role of alkali halides in realizing large area growth of transition metal-di-chalcogenide layers has been validated, the film-growth kinematics has not yet been fully established. This work presents a systematic analysis of the $MoS_2$ morphology for films grown under various pre-treatment conditions of the substrate with sodium chloride (NaCl). At an optimum NaCl concentration, the domain size of the monolayer increased by almost two orders of magnitude compared to alkali-free growth of $MoS_2$. The results show an inverse relationship between fractal dimension and areal coverage of the substrate with monolayers and multi-layers, respectively. Using the Fact-Sage software, the role of NaCl in determining the partial pressures of Mo- and S-based compounds in gaseous phase at the growth temperature is elucidated. The presence of alkali salts is shown to affect the domain size and film morphology by affecting the Mo and S partial pressures. Compared to alkali-free synthesis under the same growth conditions, $MoS_2$ film growth assisted by NaCl results in ≈ 81% of the substrate covered by monolayers. Under ideal growth conditions, at an optimum NaCl concentration, nucleation was suppressed, and domains enlarged, resulting in large area growth of $MoS_2$ monolayers. The monolayers were found to be free of unintentional doping with alkali metal and halogen atoms and exhibit high crystallinity and excellent opto-electronic quality.

## Introduction

Two dimensional (2D) materials possess a layered atomic structure[1–3] and exhibit unique properties of light emission, anomalous lattice vibrations, valley polarization, excitonic dark states and strong light-matter interactions.[4–8] The interest in $MoS_2$ is largely triggered by the fact that its energy band gap becomes direct at the monolayer limit ($E_g$ = 1.85 eV), which makes it promising for electronic and opto-electronic applications.[9–11] In order to successfully exploit the unique attributes of 2D-$MoS_2$ and to integrate them into existing application platforms, it is essential to critically assess the growth recipes and evaluate both qualitatively and quantitatively the quality of the as-grown films.

Recent work has shown that by modifying the growth recipes it is possible to grow wafersize $MoS_2$ films with various morphologies, such as compact triangles, star-like flakes, and dendrites.[8,12–19] In order to obtain a specific morphology, there are pronounced differences in the growth processes, such as the use of $O_2$ to enable monolayer growth[20], ambient or low pressure for star-like growth[17,18,21–23] and the use of metal-organic precursors for largearea growth.[24] Tuning these morphologies in a simple experimental setup with reproducible results is still a challenge and is one of the objectives of this work. Recent modifications to the growth approach has been the addition of alkali halides like sodium chloride (NaCl) to promote 2D film growth, resulting in monolayers and multi-layers of $MoS_2$.[24–27] The films were found to have markedly different morphologies than those grown without NaCl. The ambiguity in these results in terms of the observed morphology change and the possible mechanisms that account for this change remain an issue that warrants further investigation. Another challenge that impedes establishment of a reproducible recipe is the post-growth film analysis and quantification of the complex film pattern. There are reports of a conventional technique of fractal dimension analysis applied on µm-scale islands of $MoS_2$.[13,17] In a recent work, Nie et al.,[28] demonstrated the use of kinetic Monte Carlo simulation coupled with first principles calculations to study a specific case of monolayer deposition. No such work has been attempted on large area (cm scale) $MoS_2$ films. Such a study is essential because a quantified film complexity (fractal dimension) can be correlated to the crystallization process, that would provide insights into the growth mechanism.

In this work, we report on a reproducible experimental set-up to achieve controlled film growth of $MoS_2$ monolayers and multi-layers using an atmospheric pressure CVD system. The two major highlights of the growth method are: (i) pre-treating the growth substrate using a NaCl mixture, and (ii) pre-depositing the Mo precursor directly onto the substrate prior to its loading into the CVD system. The advantage of the proposed strategy lies in the fact that it is possible to tune the growth regime by varying the NaCl concentration on the pre-treated growth substrate. At low NaCl concentrations, homogeneously distributed $MoS_2$ monolayers in the form of triangles or three-point stars were obtained with controllable dimensions. At high NaCl concentrations, continuous $MoS_2$ films with multi-layered domains were formed. At an optimum concentration of NaCl, the areal coverage of monolayers peaked at ≈ 81%.

The morphology, physical dimensions, stoichiometry and crystal quality of the $MoS_2$ films were investigated using optical, scanning electron and atomic force microscopes. The optical quality of the as-grown films was also assessed by means of Raman and photoluminescence (PL) spectroscopy. For better image clarity, contrast imaging was enabled by color threshold, which is a simple technique of segmenting images. The fractal dimensions of monolayers and multi-layers was subsequently determined using a conventional box-counting method. Various morphologies of monolayers and multi-layers with shapes ranging from compact triangles, star-like flakes, circular dendrites, to non-symmetrical fractals covered a 2×2 $cm^2$ substrate area. Using contrast imaging, (black/white) the evolution of the morphology trend was assessed qualitatively, and the fractal dimensions are determined for a quantitative assessment. Finally, using the thermochemical software FactSage, we relate the evolution of the various morphologies to the time-dependent environment of the precursors in terms of the partial pressures of the gaseous reactants in the growth chamber.

## Experiment

*$MoS_2$ growth procedure.* 2D $MoS_2$ crystals were synthesized by a complex reaction involving NaCl, $MoO_3$ and sulfur vapors in atmospheric pressure and at high temperature, in an inert Argon environment. Two different substrates referred to as the precursor and growth substrates were used in this experiment. An $Si/SiO_2$ substrate (2×0.7 $cm^2$) was spin coated at 3000 rpm for 1 min with an aqueous solution of 20 mg·$mL^{-1}$ $MoO_3/NH_3·H_2O$ (Molybdenum(VI) oxide, ≥99.5%, Ammonia solution 28-30%, MilliporeSigma). This substrate functions as the Mo precursor source, thereby allowing controlled concentration of $MoO_3$ vapors in the growth chamber. A second $Si/SiO_2$ substrate (2×2 $cm^2$) that functions as the growth substrate was prepared by spin coating (3000 rpm for 1 min) an aqueous solution of $NaCl/NH_3·H_2O$ (Sodium Chloride, ≥99.5%, MilliporeSigma) onto its surface. These pre-treated precursor and growth substrates were then placed over an alumina boat, at the center of the high temperature zone of a quartz tube (2-inch diameter) in a 3-zone Carbolite furnace. A series of pre-treated samples with different $NaCl/NH_3·H_2O$ concentrations, namely 0.25 mg·$mL^{-1}$, 0.5 mg·$mL^{-1}$, 1 mg·$mL^{-1}$, 2 mg·$mL^{-1}$, 4 mg·$mL^{-1}$, and 5 mg·$mL^{-1}$ were prepared. Two similar pre-treated substrates were placed on either side of the precursor substrate to enable a study of the carrier gas dynamics on the film morphology. A separate alumina boat containing 30 mg of sulfur powder (≥99%, MilliporeSigma) was placed in the independently controlled low temperature zone, located near the end of the quartz tube.

Figure 1(a) shows the corresponding locations of the precursor and growth substrates along with the sulfur source in the three-temperature-zone CVD system. Prior to growth, the tube was purged with argon at an estimated flow rate of 500 sccm. The growth temperature of 800 °C was reached in the main zone of the furnace at a ramp rate of 20 °Cmin$^{-1}$ and held constant for 10 min. During this time period, the argon flow rate was maintained at 70 sccm. The temperature of the sulfur powder located at the edge of the secondary zone of the furnace was about 120 °C, well above the melting point of sulfur. Sulfur vapors were carried by the inert gas flow towards the main zone of the furnace where it reacts with the Mo precursor. After 10 min, the furnace was gradually cooled to room temperature.[18]

*Characterization Details.* An Olympus microscope with a maximum magnification of 1000 × was used to capture images of the MoS$_2$ films. AFM images were taken using the tapping mode of NanoSurf Mobile S AFM. Raman and photoluminescence spectra were collected from a Nicolet Almega XR Dispersive Raman spectrometer equipped with a 532 nm laser and a TE-cooled silicon CCD array detector with high-resolution gratings. The system is equipped with an Olympus BX51 microscope with motorized stage and spatial resolution down to 1 $\mu$m. The spectrometer was used in the backscattering geometry, providing a spectral resolution of approximately 1.0 cm$^{-1}$ per CCD pixel element. For PL, excitation from the 532 nm laser was focused on the sample using a 100 × objective lens. Scanning electron microscopy (SEM) was performed in field emission mode using a Hitachi SU1510 SEM operating at 10 kV. Energy dispersive X-ray (EDX) spectroscopy was performed on the same system at an accelerating voltage of 20 kV using a Bruker Quanta70 X-ray acquisition system with a Si drift detector. For electron transport measurements an MoS$_2$-based backgate field effect transistor (FET) was fabricated. Isolated MoS$_2$ monolayers were transferred onto a SiO$_2$ (285 nm dry oxide)/ p++ Si substrate, using a standard wet transfer procedure.[29] Acetone and isopropanol rinses were used to remove any chemical residue. After UV-photo lithography using the MJB3 mask aligner, Au (50 nm) contacts were deposited via thermal evaporation to form the source-drain (S/D) contacts. The highly doped Si substrate served as the back gate of the FET. Electrical measurements to characterize the fabricated device were made using the Agilent B1500 semiconductor parametric analyzer.

## Results and Discussion

The morphometric study unveiled a correlation between the NaCl concentration during the substrate pre-treatment phase and the subsequent coverage of the substrate with $MoS_2$ crystals (monolayers and multi-layers). Two identical growth substrates coated with 0.5 mg·mL$^{-1}$ NaCl/NH$_3$·H$_2$O were placed on either side of the source substrate which was coated with the Mo precursor of 20 mg·mL$^{-1}$ MoO$_3$/NH$_3$·H$_2$O. Figure 1(a) is a schematic illustrating the positioning of the source and growth substrates in the furnace, an arrangement that ensures a controlled supply of the Mo precursor. The experiment had two identical growth substrates in terms of their pre-treatment conditions. By keeping the amount of molybdenum (Mo) and sulfur (S) constant in each run, it was possible to achieve highly reproducible growth results. The NaCl is expected to function as a seeding promoter in the early stages of growth,[26,30] and its role in this work is discussed in detail in a following section.

### Results

As seen in Figure 1(b), the surface of the growth substrate (sample 1) that was placed at the upstream end of the gas flow shows clear optical contrast along the width of the sample. Based on this contrast, the entire surface of sample 1 was divided into two regions, shown enclosed by two boxes marked by red and yellow dashed lines. A magnified view of the boxed regions of sample 1 show that the area encompassed by each bounding box contains a uniform distribution of MoS$_2$, though their areal coverage is very different. The red bounding box region contained a low density of single, compact, triangle shaped MoS$_2$ domains with clearly visible boundaries and lateral sizes in the range of 90 - 100 $\mu$m. On the other hand, the yellow bounding box region of sample 1 contains a high density of uniformly distributed MoS$_2$ films. The observed color difference between the two regions on sample 1 is due to a difference in the areal coverage of the substrate by MoS$_2$. The lack of a color contrast in sample 2 that was at the downstream end of the flowing gas indicates that the entire surface is covered by a high density of continuous MoS$_2$ film. The entire sample 2 is therefore shown bound by the yellow box in Figure 1(b). These results indicate that the location of the growth region with respect to the carrier gas flow as well as the concentration of the precursors in the vicinity of the growth substrate affects the areal density of the growing films. At this stage, the role played by NaCl in the overall growth process was still not evident.

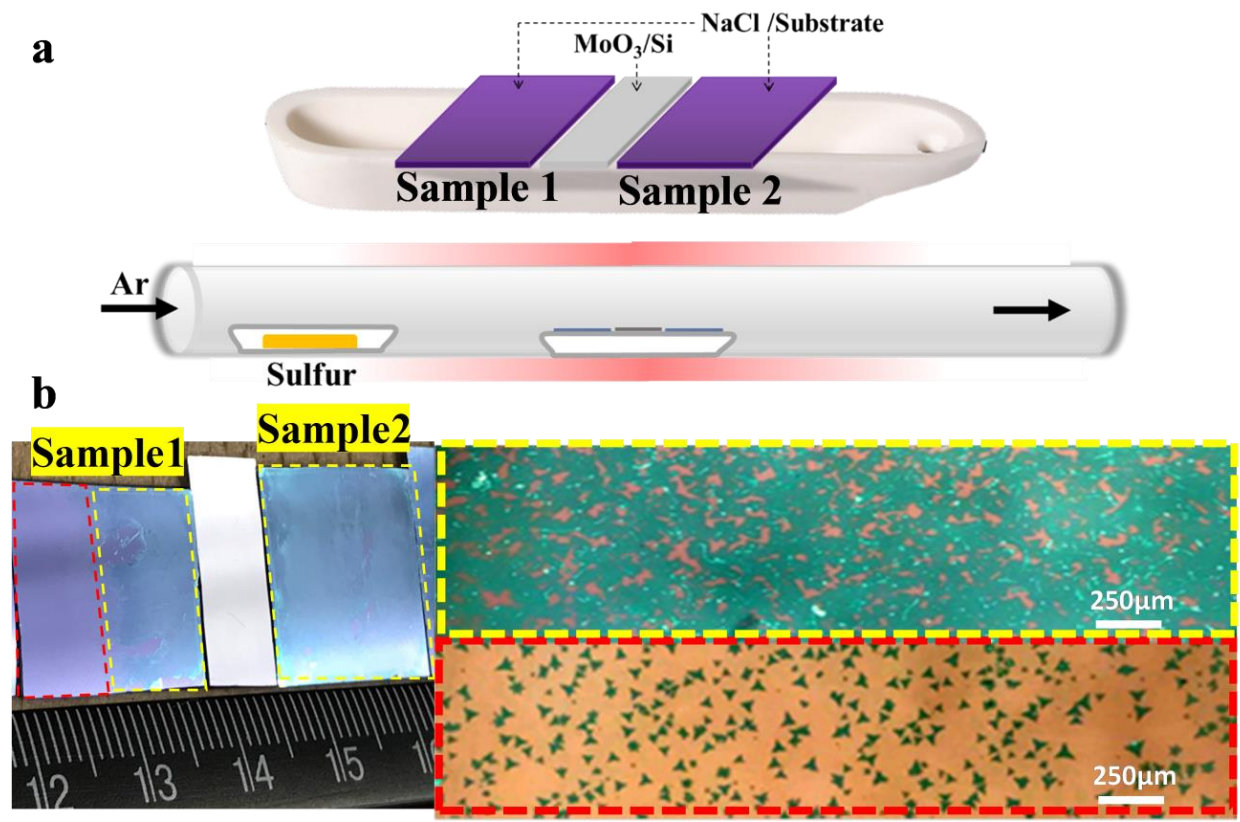

Figure 1: Large area growth of MoS$_2$. (a) Schematic of the experimental setup showing the placement of the substrates in the growth chamber, where samples 1 and 2 are coated with 0.5 mg·mL$^{-1}$ NaCl solution. (b) Optical image of the 2×2 cm$^2$ SiO$_2$/Si substrates following MoS$_2$ growth. A marked color difference on the SiO$_2$/Si substrate delineates the substrate into two regions. The dashed enclosed area in red represents the region with isolated MoS$_2$ flakes. The dashed enclosed region in yellow is the area that had continuous flakes.

To elucidate the role of NaCl in the growth process, several samples with different concentrations (0.25, 0.5, 1, 2, 4, and 5 mg·mL$^{-1}$) of NaCl/NH$_3$·H$_2$O were spin coated onto 4.0 cm$^2$ SiO$_2$/Si substrates. A pilot substrate with no NaCl pre-treatment was used as the control substrate. As shown in the optical images of Figure 2, the areal coverage of MoS$_2$ on the growth substrates increased from 5% to 100% as the NaCl concentration varied from 0 mg·mL$^{-1}$ to 2 mg·mL$^{-1}$. The pilot substrate that received no pre-treatment with NaCl was found to contain very few, randomly distributed small domains of MoS$_2$ monolayers (Figure 2(a)). At the lowest NaCl concentration of 0.25 mg·mL$^{-1}$ (Figure 2(b)), isolated and homogeneously distributed MoS$_2$ triangles were produced. However, at a NaCl concentration of 0.5 mg·mL$^{-1}$ (Figure 2(c)), the triangular flakes merged, resulting in a continuous film with randomly distributed epilayers. As the NaCl concentration was further increased to 1 mg·mL$^{-1}$ and 2 mg·mL$^{-1}$, the epilayers coalesced in some regions resulting in the formation of multi-layered MoS$_2$ films (Figure 2(d,e)). Further increase in the NaCl concentration to 4 mg·mL$^{-1}$ and 5 mg·mL$^{-1}$ resulted in a drastic reduction in the areal coverage of substrate by monolayers and multi-layers, causing more of the bare substrate to be exposed. (Figure 2(f,g))

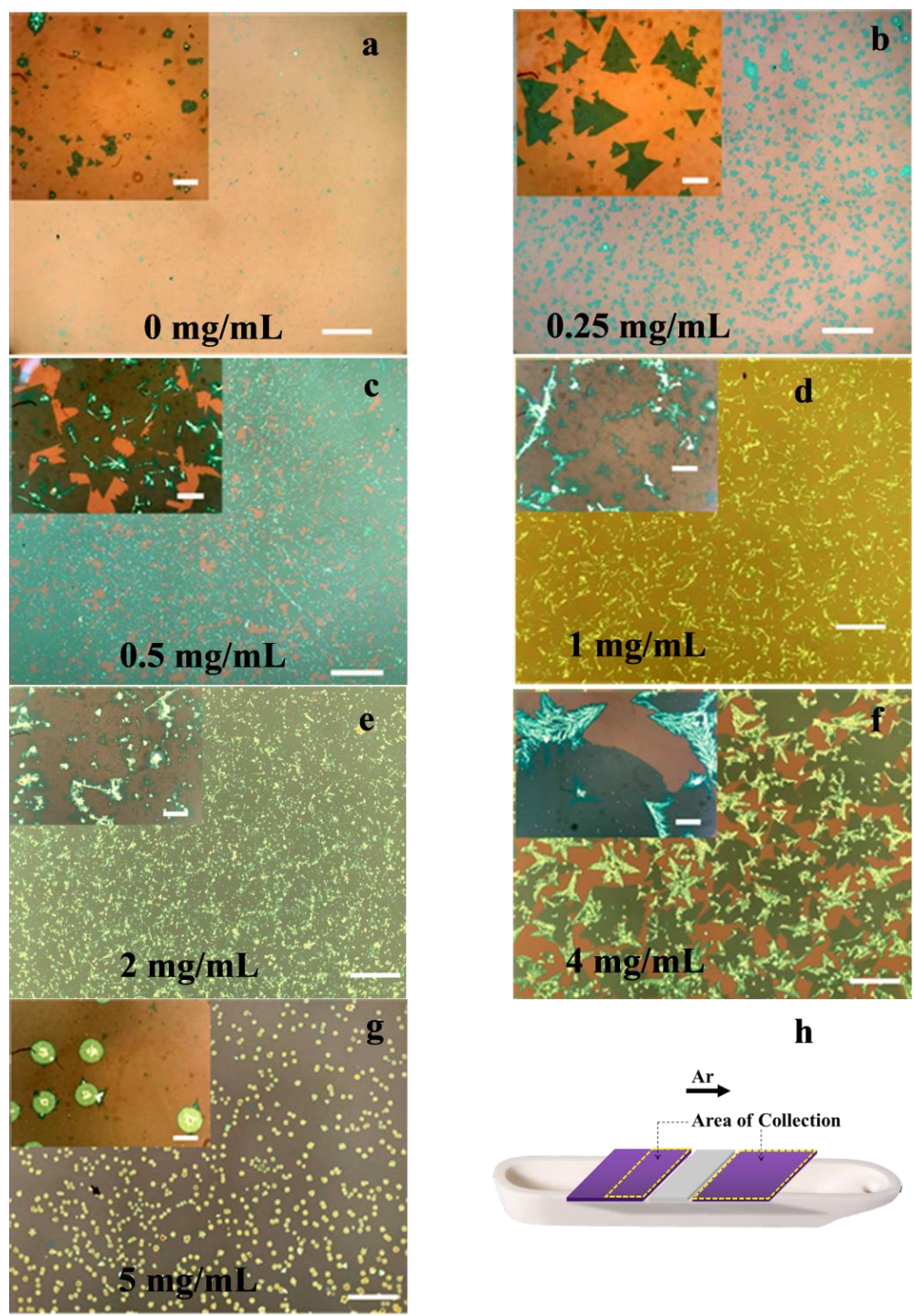

Figure 2: Optical heterogeneity in the MoS$_2$ film for different NaCl pre-treatment concentrations. Inset shows a magnified view of the morphology. (a) View of the substrate that was not pre-treated with NaCl; (b-g) Image of the substrate showing MoS$_2$ films for different NaCl pre-treatments: (b)0.25 mg·mL$^{-1}$, (c)0.5 mg·mL$^{-1}$, (d)1 mg·mL$^{-1}$, (e)2 mg·mL$^{-1}$, (f)4 mg·mL$^{-1}$ and (g)5 mg·mL$^{-1}$. Scale bars: 250 $\mu$m, inset scale bars are 25 $\mu$m. (h) Position

map showing marked areas (enclosed by dashed yellow line) used in the optical images of (a) to (g).

To compare the areal coverage of the substrate with mono- and multi-layers, the images in Figure 2 were processed using a Color Threshold module by Photoshop CC software. This technique of image analysis that uses pixels of colors to detect objects of consistent color values enables a clear determination of the $MoS_2$ film coverage. The optical contrast between monolayers, bi-layer, and tri-layers as well as the bare substrate is obvious and has been reported before.[18]

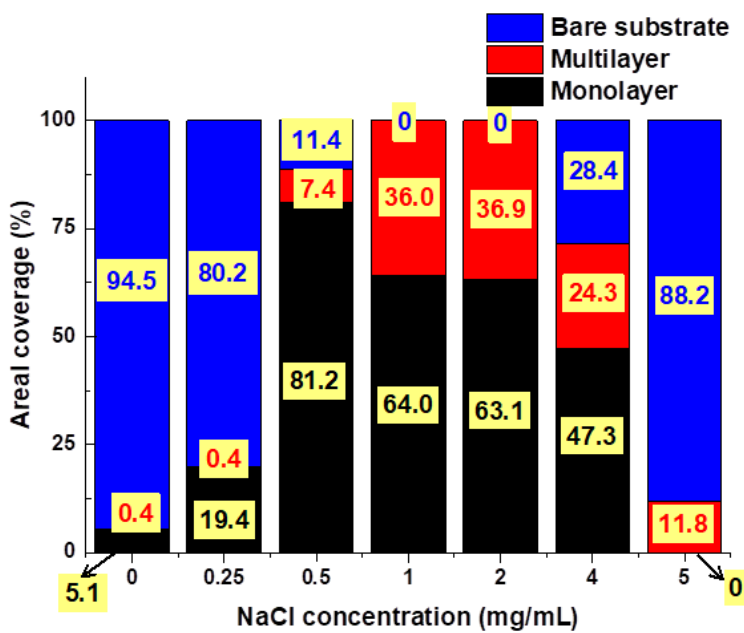

Figure 3: Plot showing the dependence of areal coverage profile of the $MoS_2$ film on NaCl concentrations. The results are based on the number of pixels of respective colors for monolayer, multi-layer and bare substrate for images (a) through (g) in Figure 2.

Figure 3 correlates the NaCl concentration during the pre-growth phase to the $MoS_2$ film coverage profile. As seen in the plot, $MoS_2$ monolayer coverage peaks at an NaCl concentration of 0.5 mg·mL$^{-1}$. At NaCl concentrations exceeding this value, there were multi-layer formations, for which the coverage peaked at 2 mg·mL$^{-1}$ concentration. These results show that the domain size, film thickness as well as the coverage of the substrate with monolayer and epilayer domains can be controlled by varying the NaCl concentration during the substrate pre-treatment phase. This growth recipe is promising because at low NaCl concentrations, it becomes possible to obtain isolated and homogeneously distributed $MoS_2$ monolayers with controllable dimensions, while for higher concentrations both mono-layered and multi-layered $MoS_2$ domains are formed.

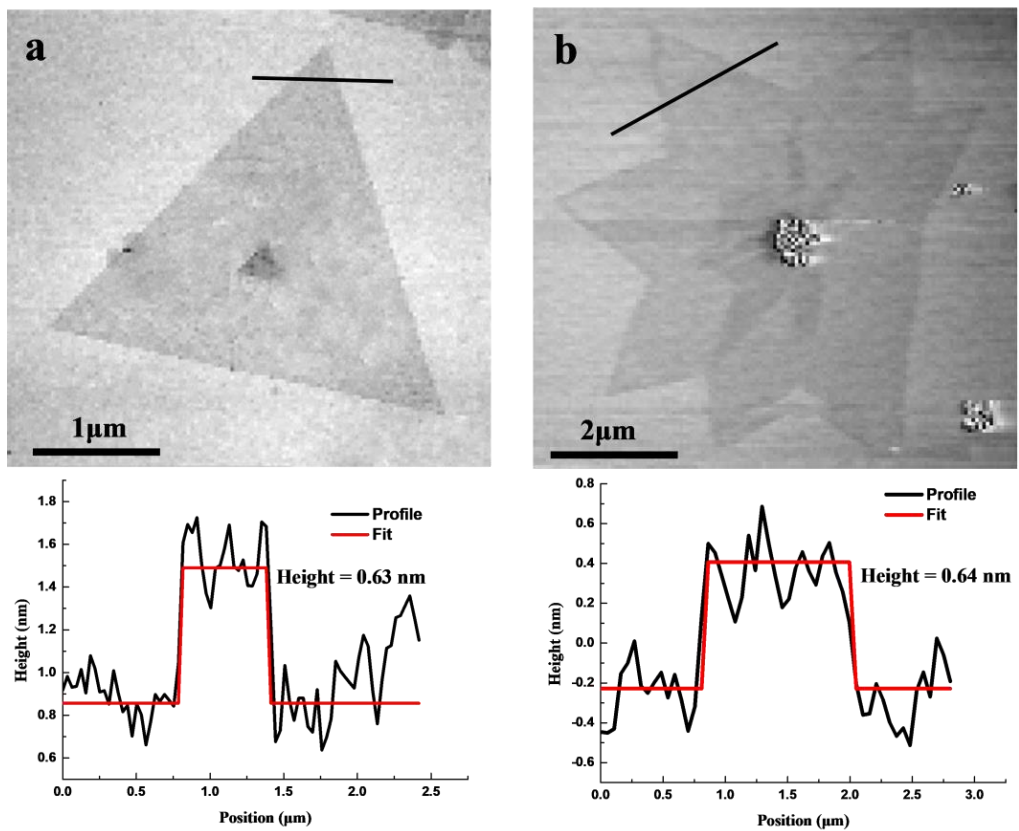

Figure 4: AFM image and corresponding height profile of $MoS_2$ domain. (a) A single triangular domain with its corresponding height profile suggesting the thickness of the $MoS_2$ film is 0.63 nm; (b) A star-like $MoS_2$ flake with the height profile extracted along the line shown in the image suggesting a thickness of 0.64 nm.

The physical characteristics of the $MoS_2$ films on the same substrate were further examined using AFM. The AFM images (Figure 4) depicts a height profile around a single isolated triangular domain and around a star-like flake. In Figure 4(a), the triangular domain is a monolayer with a few multi-layers in the center. A second morphology observed on the same substrate is shown in Figure 4(b), where the monolayer extends out spanning the region around the star-like flake in Figure 4(b). The black line that crosses the flakes, corresponding to monolayer height profiles of 0.63 and 0.64 nm, respectively, indicates the AFM sampling region.[31] The growth substrate that was pre-treated with 0.5 mg·mL$^{-1}$ NaCl/NH$_3$·H$_2$O had maximum coverage of monolayers and so was subjected to more qualitative and quantitative analysis.

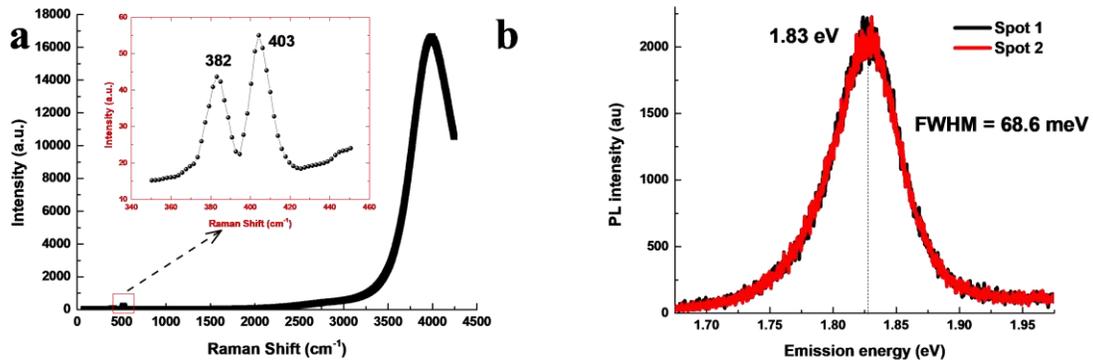

Figure 5: Results of Raman and photoluminescence spectroscopy. (a) Raman spectrum collected over a wide wavelength range to include the phonon vibration modes and the emission spectrum. Inset shows the Raman spectrum with signature double peaks of MoS$_2$. The double peak difference is indicative of the specific layer number of 2D MoS$_2$ (monolayers have $\delta k$ = 21 cm$^{-1}$). (b) photoluminescence spectrum shows strong band-edge emission peak at 1.83 eV, with a full width at half maximum (FWHM) estimated as 68.6 meV.

The Raman spectrum obtained from the 0.5 mg·mL$^{-1}$ NaCl/NH$_3$·H$_2$O pre-treated sample (Figure 5(a)) exhibits the characteristic Raman peaks of MoS$_2$ and the band-edge related emission peak at 4000 cm$^{-1}$ (corresponding to an emission wavelength of 676 nm). The double peak with the $E^1_{2g}$ mode at 382 cm$^{-1}$ and the A$_{1g}$ mode at 403 cm$^{-1}$ corresponds to a frequency difference of 21 cm$^{-1}$, characteristic of the monolayer quality of MoS$_2$.[5,18] Figure 5(b) shows the PL spectra of the same MoS$_2$ monolayer samples that were previously characterized by Raman spectroscopy. The spectrum shows an intense PL peak at around 1.83 eV with a full width at half maximum (FWHM) of about 68.6 meV. These values are consistent with the emission peak at 4000 cm$^{-1}$ measured in the Raman spectrum and also with the direct band gap nature of MoS$_2$ monolayers. Moreover, the strong PL intensity and a narrow FWHM is indicative of high crystalline quality of the film.[5]

To evaluate the electronic quality of the MoS$_2$ monolayers, back-gated field-effect transistors (FETs) were fabricated by isolating a single MoS$_2$ monolayer and patterning four electrodes on it. The transfer characteristics of the FETs obtained at drain voltages ($V_{ds}$) ranging from 0.5 to 1.5 V with a 500-mV step exhibited typical n-type behavior (Figure 6(a)). Inset shows the semi-log plots of the transfer characteristics. The output curve of Figure 6(b) shows that MoS$_2$ monolayers contacted by Au electrodes exhibits linear ohmic contact behavior in the gate range from -5 V to + 20 V.

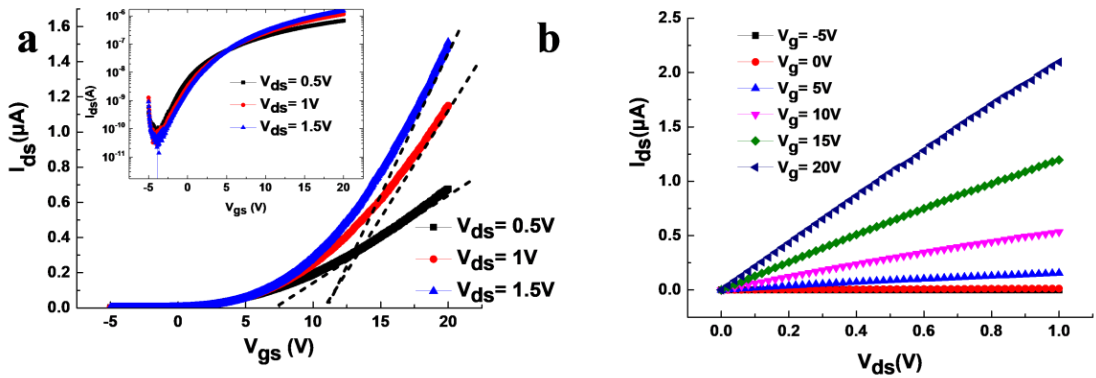

Figure 6: Evaluation of the electrical quality of the MoS$_2$-FET. (a) Gate characteristics of the device at drain source voltages of 0.5V, 1V and 1.5V. Inset is the semi-log plot of the gate characteristics. (b) I-V characteristics of the device at gate bias ranging from -5 V to +20 V.

At $V_{ds}$ = 1.0 V, the FET had a threshold voltage ($V_{th}$) of ≈11.0 V. Due to the relatively large band gap of MoS$_2$ monolayer and its atomically thin channel, a relatively high $I_{on}/I_{off}$ of $10^5$ was achieved. Considering that the gate oxide used in this work was 285 nm thick, there is room for further improvement of the $I_{on}/I_{off}$ ratio. The field-effect mobility of MoS$_2$ was determined to be 6.45 cm$^2$V$^{-1}$s$^{-1}$ which was obtained using the equation: $\mu_{FE} = \frac{L}{WCV_{ds}}\left(\frac{dI_{ds}}{dV_{gs}}\right)$, where d$I_{ds}$/d$V_{gs}$ is the transconductance extracted from the slope of the black dashed line in Figure 6(a) at $V_{ds}$ = 1V. The channel length (L) and width (W) are 4.0 μm and 6.0 μm respectively. $C_{ox}$ is the back-gate capacitance per unit area estimated to be 1.23 x 10$^{-4}$ Fm$^{-2}$ for the 285 nm-thick SiO$_2$. The carrier concentration was estimated to be 6.9×10$^{11}$ cm$^{-2}$, obtained using the equation: n$_{2D}$=C$_{bg}$ (V$_{bg}$-V$_{th}$), where C$_{bg}$, V$_{bg}$, and V$_{th}$ are 1.23×10$^{-4}$ Fm$^{-2}$, 20.0 V and 11.0 V, respectively.

Considering the low value of $I_{ds}$ at zero gate bias, a four probe measurement[32,33] was performed on the device (optical image shown in Figure 7(a)) to determine the contact resistance.

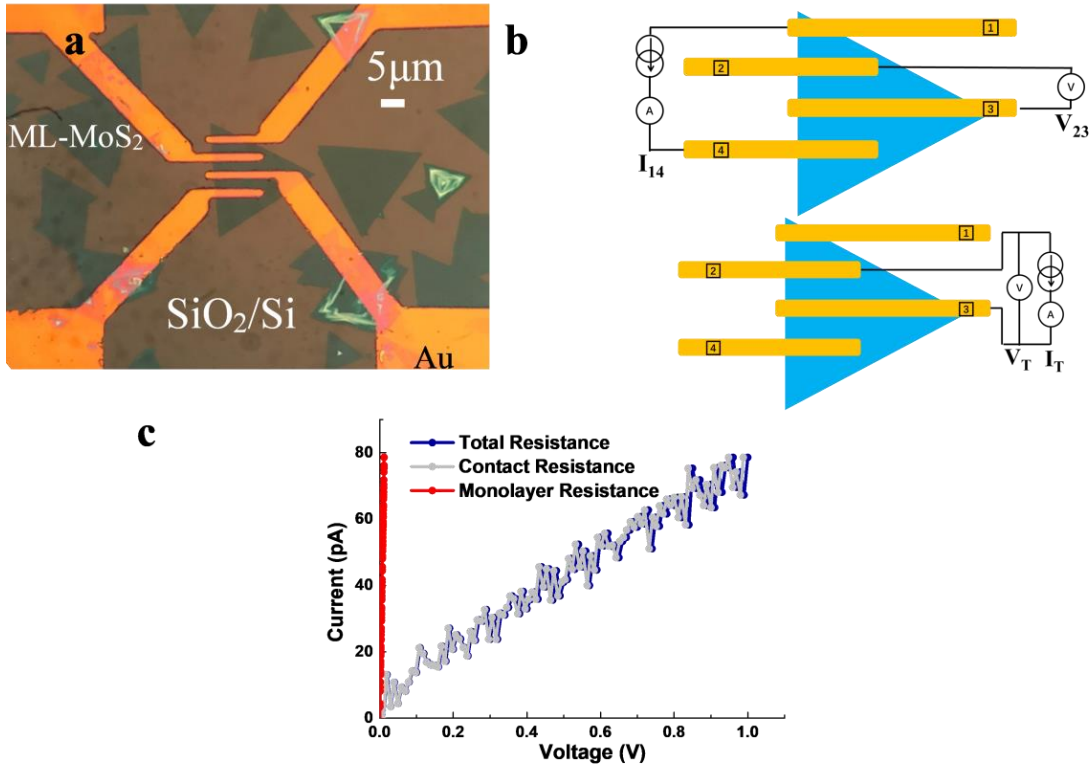

Figure 7: Device description and results of 4-terminal measurements on MoS$_2$ monolayer. (a) Optical image of the fabricated device showing a monolayer MoS$_2$ contacted by four Au electrodes. (b) Schematic of the four-probe measurement setup. Measurements of voltages and currents V$_{23}$, I$_{14}$, I$_T$, and V$_T$ are denoted in the respective circuits. The Au pads are numbered to indicate the pads between which measurements are made. (c) Current-Voltage plot for determination of monolayer and contact resistance.

A schematic of the measurement platform is shown in Figure 7(b). The film resistance was determined using the equation: $R_m = V_{23}/I_{14}$, estimated to be ≈150 MΩ. The contact resistance is subsequently obtained from the equation $V_T = I_T R_C + I_T R_m$ and is estimated to be ≈10 GΩ. It is therefore evident that the contact resistance dominates the total resistance and is the most likely reason for the reduced value of field effect mobility and drain-source current at zero gate bias. Challenges related to contacts will be addressed in a future work where experiments with improved metal-film contacts are being designed.

## Discussion

Pre-treatment of the growth substrate with NaCl and achievement of precise control over precursor (Mo-source) feed in a sulfur-rich ambient were major factors favoring the growth of uniform monolayer $MoS_2$ films with very large domain sizes (2 cm). This technique is easily scalable to enable large areal coverage on 4-inch wafers; a feat that can be achieved by increasing the size of the growth chamber. The role of NaCl in the growth process is elucidated in greater detail in the following paragraph. The texture and quality of the substrate surface following spin coating with $NaCl/NH_3·H_2O$ during the substrate pre-treatment stage is a key factor influencing $MoS_2$ film growth. During the spin-coating process, the NaCl salt crystals react with ambient moisture in the air and re-crystallizes, the process occurring in millisecond time scale.[34,35] It is these re-crystallized salt crystals that determine the texture and quality of the pre-treated substrates. An AFM image characterizing the sample surface after NaCl spin coating is shown in Figure 8. It is worth noting that since the imaging was done in ambient air and since the measurements take time to perform, the texture of the sample surface evolves with time during characterization and sample transportation.

In Figure 8, the surface quality of two SiO$_2$/Si substrates pre-treated with 0.25 mg·mL$^{-1}$ and 0.5 mg·mL$^{-1}$ NaCl solution were examined using AFM and an optical microscope (image inset). The images confirm that there is an obvious and expected difference in the areal coverage of NaCl on the substrates. The corresponding growth results are shown to the right side of the AFM images in Figure 8. As expected, a lower level of complexity in the NaCl texture results in low areal coverage of the substrate by the MoS$_2$ film. On the other hand, at a NaCl concentration of 0.5 mg·mL$^{-1}$, the substrate surface texture is complex and MoS$_2$ films cover a high percentage of the substrate. Figure 9(a) is an SEM image of highly dense MoS$_2$ monolayer grown using the 0.5 mg·mL$^{-1}$ NaCl pre-treatment. An enlarged view of a select area is shown as an inset in the right corner, with the two circles indicating the substrate treated with 0.25 mg·mL⁻¹ NaCl/NH₃·H₂O. Inset is the optical view of the same substrate. Right: Optical image of substrate, post-growth, showing sparse coverage by MoS₂. area used for composition analysis.

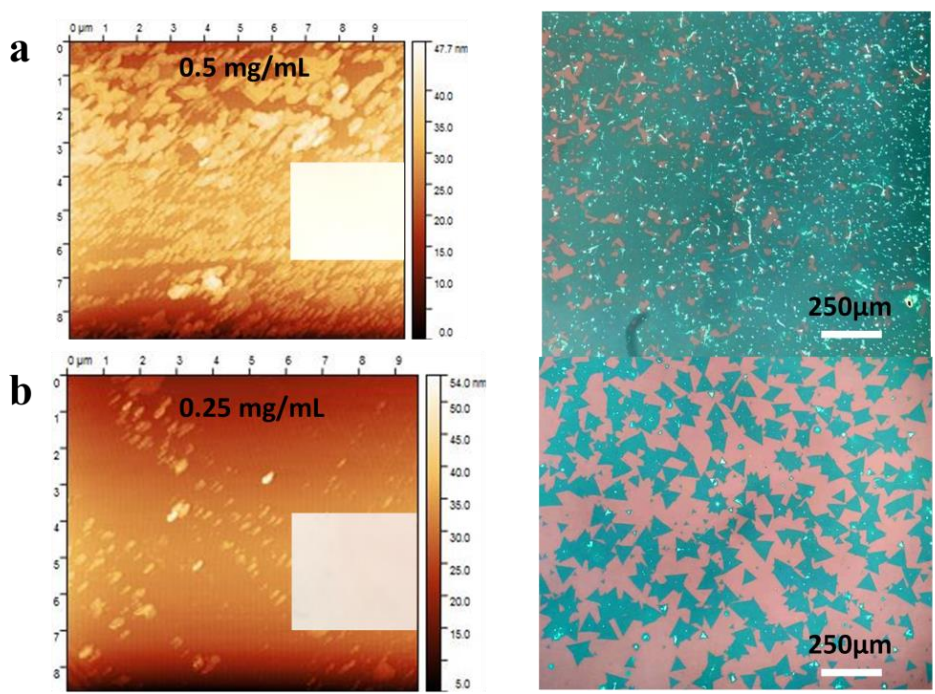

Figure 8: Distribution of monolayer MoS$_2$ on growth substrate following pre-treatment with 0.25 and 0.5 mg·mL$^{-1}$ of NaCl. (a)Left: AFM image of SiO$_2$/Si substrate treated with 0.5 mg·mL$^{-1}$ NaCl/NH$_3$·H$_2$O. Inset is the optical view of the same substrate. Right: Optical image of substrate, post-growth, showing dense coverage by MoS$_2$. (b)Left: AFM image of SiO$_2$/Si

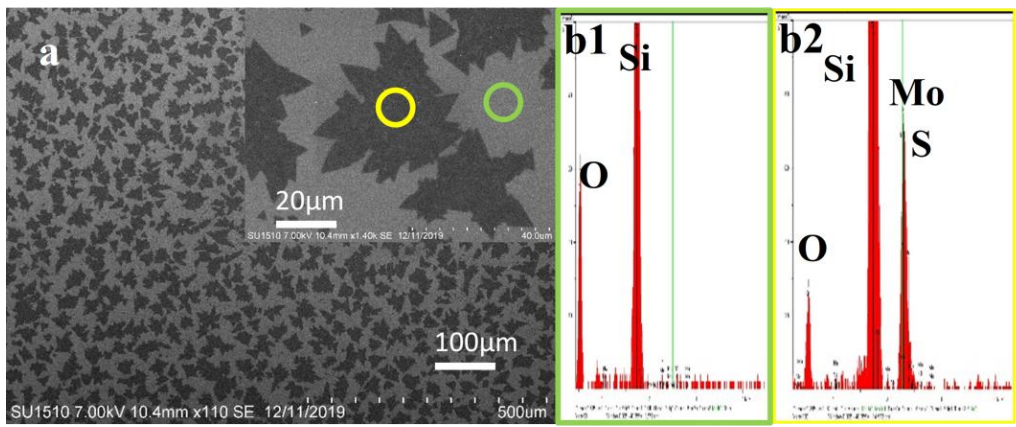

Figure 9: SEM image and EDX spectra of the MoS₂ films grown using 0.5 mg·mL⁻¹ NaCl. (a) SEM image of dendrites of MoS₂. Inset image is a magnified view of the MoS₂ film. (b1) and (b2) are the EDX spectra collected from regions indicated by the green and yellow circle in the inset of Figure (a).

An EDX analysis, performed to confirm the chemical composition of the MoS₂ crystals, indicate concentrations of Mo and S, in 9.1 wt% and 6.0 wt%, respectively. This corresponds to Mo:S ratio of 0.506, representing a MoS₂ composition close to stoichiometry. The higher Mo concentration indicates presence of sulfur vacancies which has been diagnosed as the major defect in CVD grown MoS₂ samples.[16,20–22,36–38] No signal corresponding to Na or Cl was detected. As seen in Figure 9(b1), the bare substrates show signal from the SiO₂ surface.

To quantify the various MoS₂ morphologies that were observed when using different NaCl pre-treatment, the fractal dimensions of these films were calculated using a conventional boxcounting method. The monolayer and multi-layer contrast images in Figure 10 are extracted from their corresponding images in Figure 2, corresponding to different NaCl concentrations. Figure 10 (a1) and (a2) correspond to the color contrast image of Figure 2 (a) grown from a NaCl concentration of 0 mg·mL⁻¹. Similarly, Figure 10 (b1) and (b2) are extracted from Figure 2(b) and correspond to an NaCl concentration of 0.25 mg·mL$^{-1}$. It should be noted that the image (a1), (b1), (c1), (d1), (e1) to (f1) in Figure 10 is a depiction of monolayer contrast imaging where only the monolayers are shown in white against a black background.

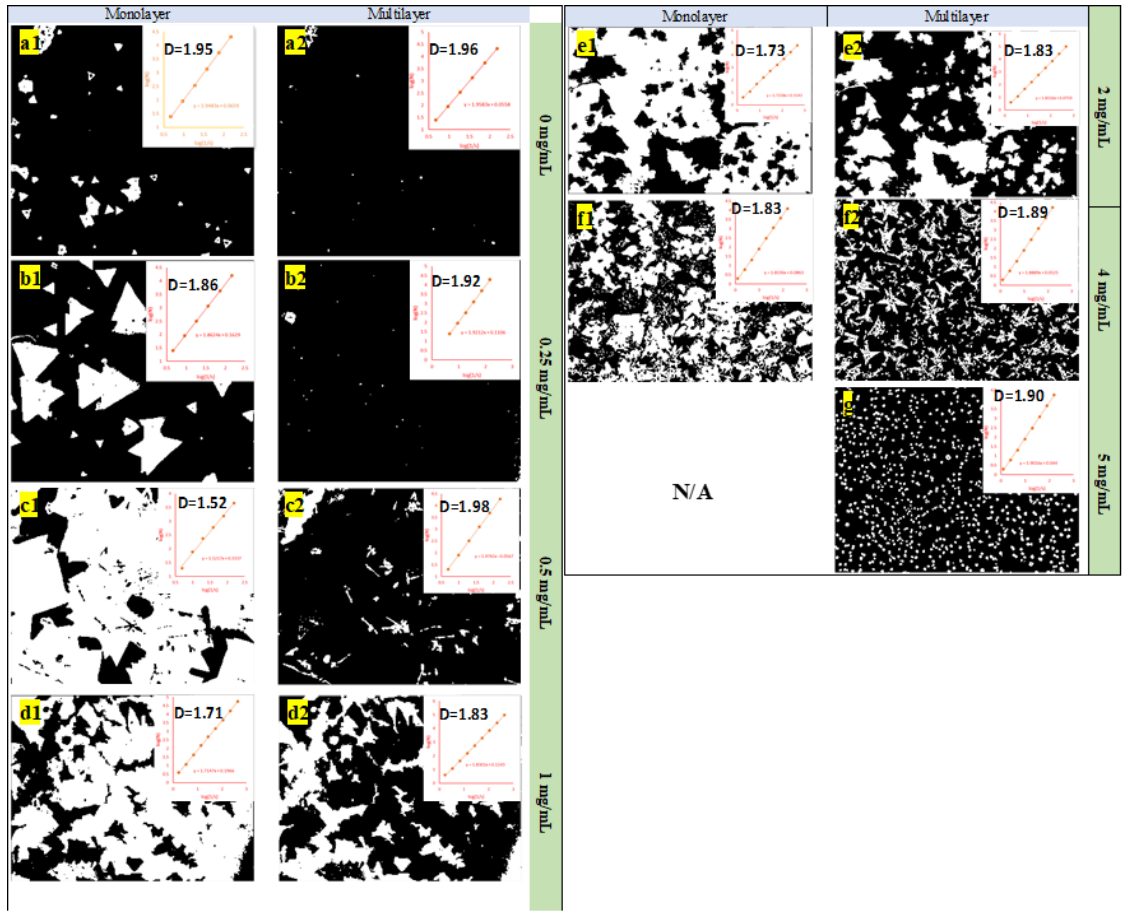

Figure 10: Contrast images of monolayer and multi-layer growth results of MoS$_2$ retrieved from the same set of images in Figure 2. (a1) and (a2) are retrieved from Figure 2(a). (b1) and (b2) are retrieved from Figure 2(b), and so on. The analysis plot of fractal dimension are shown in inset of each contrast images.

Similarly, images shown in Figure 10 (a2), (b2), (c2), (d2), (e2), (f2) and (g) represent multilayer contrast imaging with the multi-layers shown in white color against a black background. The fractal dimensions of these black and white images were then computed by Fractalyse software using the box counting method. In this method, the images were broken into smaller box-shaped pieces and each box was then analyzed to calculate the number of boxes (N) that enclosed the prescribed patterns.[13] A plot of log(N) versus log(1/s) allows for determination of fractal dimension (D) based on the equation: $\log(N) = D \cdot \log(1/s)$

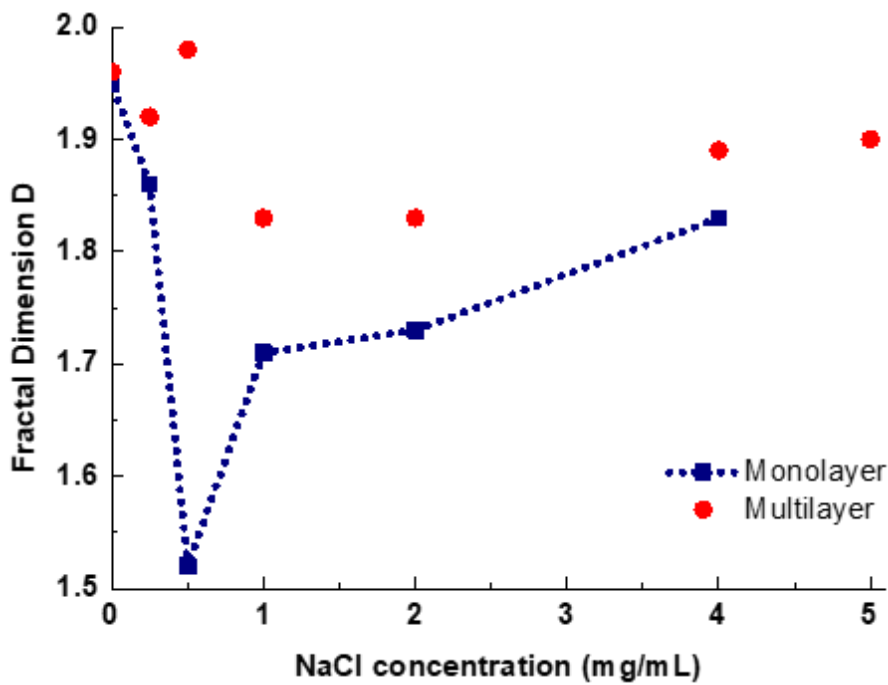

Figure 11: Analysis of the variation of fractal dimension with NaCl concentration. The data for monolayers depict a characteristic trend with a minimum value of D at NaCl concentration of 0.5 mg·mL$^{-1}$.

To determine a trend in the fractal dimension variation with NaCl concentration, a plot describing the variation of D with NaCl concentration is shown in Figure 11. As the concentration of NaCl in the pre-treatment stage is increased, the fractal dimension of the monolayer initially decreases and subsequently increases above a critical NaCl concentration. On the other hand, there is no noticeable trend in the fractal dimension of multi-layers, which appears to hold a relatively stable value. The significance of fractal dimension is that its value denotes the complexity of a pattern, a fact that is used in this work to describe the nucleation and growth mechanism of $MoS_2$ monolayers and multi-layers.[39] It is evident that the complexity of monolayer and multi-layer patterns over a mm scale (the size of images in Figure 2) relates to the areal coverage of the pattern. Although single crystal domains for different NaCl concentrations may have different shapes and possess a different D value, the cm scale of the pattern consists of cross-linked domains that influences the domain edges in the overall MoS$_2$ film pattern. Correlating the areal coverage statistics to the fractal dimension value (D), an inverse relationship is experimentally validated (Figure 3 and 11) between D and the areal coverage of the substrates by monolayers and multi-layers. This result is useful for future analysis as it can be exploited to provide insight into the growth modes of NaCl-assisted 2D MoS$_2$ films.

The FactSage software was used to determine the partial pressures of Mo- and S-based compounds in gaseous phase at 800 °C. Based on the principle of Gibbs energy minimization during equilibrium reaction, the software determines the various chemical phases present in the growth chamber as a function of time. The concentration of these phases is critically dependent on experimental parameters like the amount of S, MoO$_3$ and NaCl used in the experiment as well as the growth temperature, duration, and carrier gas flux. Figure 12(a), (b) describes this environment in terms of the partial pressures of various precursor components during the 10 min growth time with an Ar gas flow of 70 sccm. Figure 12(a) represents the condition when the substrates were not pre-treated with NaCl. As seen in this plot, sulfur-based gaseous components dominate the environment after the first couple of minutes. The higher and lower oxidation states of Mo have a significantly lower partial pressure. There is a clear difference in the growth chamber environment when NaCl is introduced into the analysis. As seen in the plot (Figure 12(b)), after the first couple of mins, there is a high concentration of MoO$_2$Cl$_2$ which dominates other oxidation states of Mo by about two orders of magnitude. Hence, based on the thermodynamics of the growth process, there is an obvious difference between the two growth cases (one without

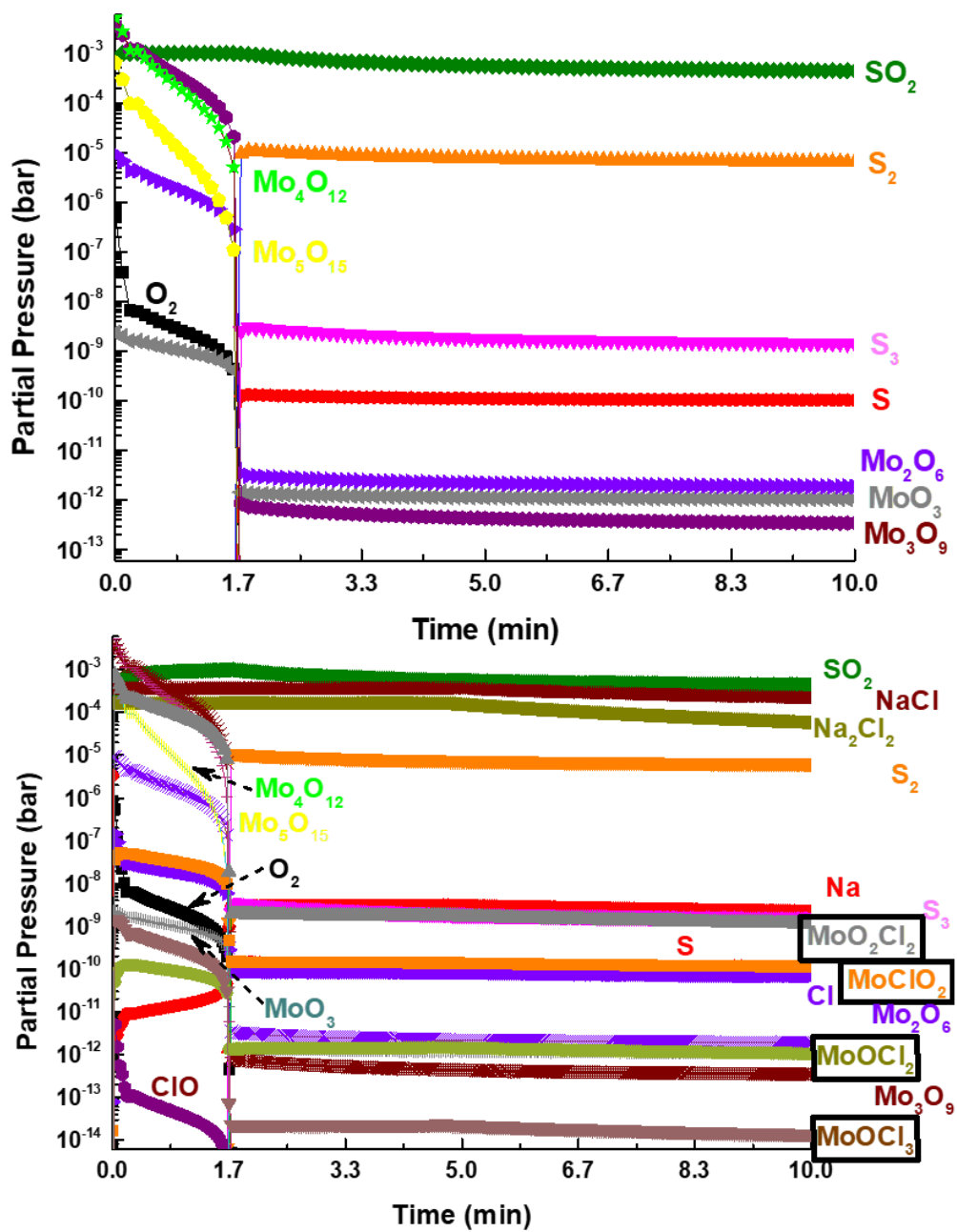

Figure 12: Thermal equilibrium calculations of partial pressure of gaseous compounds and molar amount of stable condensed phases for the growth recipe used in this work. Temperature: 800 °C, Argon gas flow rate: 70 sccm, duration: 10 minutes. (a) Plot of logarithm of partial pressure versus reaction time for $MoS_2$ film growth without NaCl. (b) Plot of logarithm of partial pressure versus reaction time for $10^{-5}$ moles NaCl (estimate corresponding to NaCl concentration of 0.5 mg·mL$^{-1}$). Compounds with less than $10^{-14}$ bar partial pressure are not presented. The highlighted compounds, $MoO_2Cl_2$, $MoClO_2$, $MoOCl_2$, and $MoOCl_3$ were attributed to the increased partial pressure of Mo species due to NaCl.

Compound names are listed besides each trend line with the same color of line. Colors of compounds are consistent for all plots.

and one with NaCl). In the presence of NaCl, in addition to the Mo-oxides there are several chloride-based Mo compounds like $MoO_2Cl_2$, $MoOCl_3$ and $MoOCl_2$. These compounds are known to be catalysts for forming various oxidation states of Mo, effectively increasing the concentration of Mo ions.[40] Theoretical calculations have shown that NaCl performs the role of a molecular seeding promoter, where it lowers the potential barrier for $MoS_2$ nucleation, thereby facilitating growth. It also facilitates layered growth by increasing the surface adhesion energy. A recent theoretical work has proposed a mechanism explaining the evolution of various $MoS_2$ morphologies as a function of incident flux, temperature and precursor ratio.[28] Here, we highlight the significance of the chalcogen to metal (X/M) ratio and relate it to the formation of various experimentally observed morphologies. According to the principles of crystal growth, there are two major competing mechanisms: (i) diffusion of atoms along the domain boundary and (ii) attachment of atoms to the domain edges. The competition between these two mechanisms result in different morphologies such as compact triangles, star-shapes, dendrites, and non-symmetrical fractals. The X/M ratio is one of the factors that governs the rate of diffusion and attachment. During the growth phase, substrates with a high concentration of NaCl on its surface suffer a decrease in the X/M ratio, due to an increase in the concentration of metal ions (Mo-containing components in vapor phase, Figure 12(b))- promoted by the reaction between NaCl and $MoO_3$ source. For stoichiometric $MoS_2$ growth, the ideal X/M ratio is 2. When this ratio decreases from the ideal value, the rate of attachment dominates over the rate of diffusion. This leads to fractal growth. In the extreme case, for high concentrations of NaCl (concentration greater than 4 and 5 mg·mL$^{-1}$), the X/M ratio decreases even further and results in a condition close to diffusion limited aggregation (DLA).[28,41] The rate of attachment in this case is much higher than the rate of diffusion and as shown in Figures 2, 3 and 10, the substrate has a high areal coverage of multi-layers as well as a high density of dendrites visible in multi-layer contrast images. At very low NaCl concentration, the phenomenon of diffusion dominates over that of attachment and MoS$_2$ film growth is inhibited. This is validated in Figure 3, where at a NaCl concentration of 0.25 mg·mL$^{-1}$, the areal coverage of monolayers and multi-layers is low at 19.4% and 0.4% respectively. A moderate increase in the NaCl concentration to 0.5 mg·mL$^{-1}$ increased the areal coverage of the substrate with MoS$_2$ monolayers to the extent of 81.2%. In this regime, the mechanisms of diffusion and attachment both contribute to film growth.

## Conclusion

The work presented here highlights the effect of substrate pre-treatment (using NaCl and Mo-precursors) on film growth of MoS$_2$ monolayers and multi-layers. An optimization of the concentration of NaCl enabled a significant increase in the domain size of the film. At an NaCl concentration of 0.5 mg·mL$^{-1}$, approximately 81% of the substrate was covered by high-crystal quality monolayers. The film complexity as quantified by fractal dimension calculations reveal a relationship between the NaCl concentration and fractal dimensions. Using FactSage software, theoretical estimates of the partial pressure of Mo and S gaseous components were computed and used to explain the growth kinematics. The chalcogen/metal (X/M) ratios determine various crystal morphologies based on two competing mechanisms that favor diffusion or attachment of atoms along the domain boundary. At low NaCl concentrations, the ratio of X/M is high, causing diffusion to dominate over attachment, resulting in low areal coverage growth of MoS$_2$ films. On the other hand, at high NaCl concentrations, the ratio X/M is low, and attachment dominates over diffusion, resulting in fractal growth. At an optimum NaCl concentration, the effects of diffusion and attachment contribute equally towards large areal growth of MoS$_2$ layers. This work therefore analyzes the growth kinematics and provides a reproducible recipe for tuning the fractal dimension of 2D MoS$_2$ films to the order of several cms.